\documentclass[preprintnumbers,amsmath,amssymb,subfloat,floatfix,10pt,prd,superscriptaddress,nofootinbib]{revtex4}
\usepackage{graphicx,epsfig} 
\usepackage{dcolumn}
\usepackage{bm}
\usepackage{caption}
\usepackage{subfig}
\usepackage{amssymb}
\usepackage{color}
\usepackage{amsmath}
\usepackage{amsmath}
\usepackage{amsfonts}
\usepackage{amssymb}
\usepackage{graphicx}
\usepackage{dcolumn}
\usepackage{bm}

\begin{document}


\title{ Anisotropy effects on Baryogenesis in $f(R)$-Theories of Gravity }
\author{ A. Aghamohammadi}
 \email{a.aqamohamadi@gmail.com; a.aghamohamadi@iausdj.ac.ir }
 \affiliation{  Islamic Azad University,  Sanandaj Branch, Sanandaj  Iran.}
 \author{ H. Hossienkhani}
\email{Hossienhossienkhani@yahoo.com}
\affiliation{ Hamedan Branch, Islamic Azad University, Hamedan, Iran}
\author {Kh. Saaidi}
\email{ksaaidi@uok.ac.ir}
\affiliation{ Department of Physics, Faculty of Science, University of Kurdistan,  Sanandaj, Iran}

\date{\today}

\begin{abstract}
We study  the $f(R)$ theory of gravity in an anisotropic metric and its effect on the baryon number to entropy
ratio. The mechanism of gravitational baryogenesis  based on the CPT-violating gravitational interaction between  derivative of the Ricci scalar curvature and the baryon-number current is investigated in the context of  the $f(R)$ gravity. The gravitational baryogenesis  in the Bianchi type I \textbf{(BI)} {\bf Universe} is examined. We  survey   the effect of anisotropy of the {\bf Universe} on the baryon asymmetry from point of view the $f(R)$-theories of gravity and its effect on  $n_{b}/s$  for radiation dominant regime.
\end{abstract}


\maketitle

\section{Introduction}
In recent decades,  It was indicated that our {\bf Universe} is in a positive accelerating  expansion phase  \cite{1f, sp, tm}, and the standard model of gravity could not explain this phenomena. This is a shortcoming for Einsteinian theory of gravity and several attempts have been accomplished to solve it.\\
 One way for explaining this  positive accelerating expansion is modified gravity. In fact some people have believed this shortcoming of Einsteinian theory of gravity is coming from the geometrical part of Hilbert-Einstein action and try to modify it by replacing a function of Ricci scalar, $f(R)$, instead of $R$ in the Hilbert-Einstein action and so-called $f(R)$ model of gravity. The reason relies on the fact that they allow to explain, via gravitational dynamics, the observed accelerating phase of the  {\bf Universe}, without invoking exotic matter as sources of dark matter. There are another modified gravity formalism  which are completely different with $f(R)$ model of gravity.
The various aspect of $f(R)$ models of gravity is investigated in  \cite{fre, so, n6, n7,  sc2, sn1, cs, cno, amc, ak, kaa, aks, ka,kha,r1}.
{\bf The unification of dark energy and early time inflation with late time acceleration from $f(R)$ theory to all Lorentz non-invariant theories
is discussed by Nojiri and Odintsov  \cite{r2}.  In addition to this   model,   a singular analog of $R^2$ inflation is studied in \cite{r3,r4,r5,r6}.} It is well known, the most of researches about the dynamical evolution of the  {\bf Universe} have been  done in a homogenous and isotropic space-time background such as FLRW. But tiny deviation from  isotropy at the level of $10^{-5}$, has also been suggested  by Bennett \textit{et al} (1996) and afterwards this suggestion was  confirmed by high resolution WMAP data.  Although,  by considering  the present  {\bf Universe}, it is known that the anisotropy is small. The possible effects of anisotropy in the early {\bf  Universe} have investigated with  Bianchi I type (BI)  models from different point of view \cite{ko, ku, Y1, Y2, 17,18}. {\bf The  Kasner-type is a special class
of BI model, for which cosmological scale factors evolve as a power law in time. In GR the vacuum Kasner solutions \cite{r7} and their fluid filled counterparts, the
BI models, proved useful as a starting point for the investigation of the
structure of anisotropic models. Barrow and Clifton  \cite{r8,r9} have recently shown that
it is also possible to find solutions of the Kasner type for $R^n$-gravity models.}
Recently, Hossienkhani \textit{et al}. \cite{18a} discussed the effects of the anisotropy on the evolutionary behavior DE models and compare with the results of the standard FRW, $\Lambda$CDM and $w$CDM models. Also, they shown that the anisotropy is a non-zero value at the present time although it is approaching zero, i.e. the anisotropy will be very low after
inflation. \\
The main purpose  of this paper is that  show  $f(R)$-theories of gravity provide a framework in which the gravitational
baryogenesis may occur in the anisotropy  background  and  lead to  the observed baryon asymmetry in the {\bf  Universe}.
The cause  of the baryon number asymmetry is still an open problem of the particle physics and cosmology. The measurements of cosmic microwave background \cite{1}, the absence of $\gamma$ ray emission from matter- antimatter annihilation \cite{2} and the theory of  Big-Bang nucleosynthesis \cite{3,mos} imply that  there  is  matter in excess of   antimatter in the {\bf Universe}. The observational results indicates  that the ratio of the baryon number to entropy density is approximately $n_{b}/s \sim 10^{-10}$. In \cite{4}, it was pointed out that a process generating,   baryon,  antibaryons and those different rates  may be  satisfied by the  following three conditions
\begin{enumerate}
\item
  Baryon number non-conservation.
\item
C- and CP-symmetry violation.
\item
 Deviation from thermal equilibrium.
\end{enumerate}
To satisfy two later  conditions, the conventional approach has been to introduce interactions which violate C and CP in vacuo and a period in which the {\bf  Universe} is out of thermal equilibrium. Referring to Lorentz's and CPT symmetries, the more general setting in which they have been studied is the
Standard Model Extension (SME) \cite{5,odin}. However, a dynamically violation of CPT may be lead  to generation  the baryon number
asymmetry also in regime of thermal equilibrium \cite{6}.  They have introduced an interaction between Ricci scalar curvature and any  current that give rise to net $B-L$ charge in equilibrium  ($L$ is lepton number) which dynamically violates CPT
symmetry in expanding Friedmann Robertson Walker (FRW) {\bf Universe}. As a consequence, in this work, the baryon number asymmetry can not be directly generated in
radiation dominated epoch. But in \cite{7}, in the framework of modified theories of gravity, it was shown that the baryon asymmetry may be generated even in the radiation dominated era. We explicitly calculate the asymmetry in our scenario and compare it to the baryon asymmetry of the {\bf  Universe}. We follow the work of Ref. \cite{7}. Ref. \cite{7a} studied the gravitational baryogenesis scenario, generated by an $f(T)$ theory of gravity. {\bf But in  \cite{r10}, it was calculated the baryon to
entropy ratio for the Gauss-Bonnet term and by using the observational constraints.  Moreover,  in the context of modified it is possible to generalize the gravitational
baryogenesis mechanism,  in the  $f(R)$ theory of gravity \cite{r11} and the Loop Quantum Cosmology  \cite{r12}.}  In a less symmetric background spacetime, however,
some possibility of the enhancement of the baryon asymmetry was argued in \cite{7b,7c}.  They showed that the baryon asymmetry will increase with the anisotropy of the {\bf  Universe}.  In this way more general cosmological evolutions can be
considered and the resulting baryon-to-entropy ratio is compatible to the observational data.
 This paper is organized as follows. In the next section we first  review the field equation of $f(R)$ gravity theory in the BI {\bf  Universe}. In Sect. III, we study the gravitational baryogenesis in the anisotropic {\bf  Universe}. At last,  we summarize our results in last section.

\section{Description and general  properties of  the model}
\subsection {$f(R)$ Theory}
The action $S$ of $f(R)$ gravity with general matter is given by
\begin{equation}\label{1}
S=\int  \sqrt{-g}\; d^4x\left [\frac{f(R)}{2} + L_m(\psi, g_{\mu\nu}) \right],
\end{equation}
where $f(R)$ is an arbitrary function of Ricci scalar, $R$, $L_m = L_m(\psi, g_{\mu\nu})$, $\psi$,  $g_{\mu\nu}$  and $g$ are the matter Lagrangian,  the matter field,  the metric of space-time and the determinant of metric respectively.   Variation of (\ref{1}) with respect of $g^{\mu\nu}$ gives
\begin{equation}\label{2}
R_{\mu\nu}f'-\frac{1}{2}fg_{\mu\nu}+\left( g_{\mu\nu}\Box-\nabla_{\mu}\nabla_{\nu}\right)f'=\kappa^2 T^m_{\mu\nu},
\end{equation}
where prime represents  the derivative with respect to  the scalar curvature  $R$, $\Box\equiv \nabla_{\alpha}\nabla^{\alpha}$ and $T^m_{\mu\nu}$ are the covariant  d'Alembert  operator  and the stress-energy tensor of matter respectively and $T^m_{\mu\nu}$  is defined by
\begin{equation}\label{3}
T^m_{\mu\nu} = \frac{-2}{\sqrt{-g}}\frac{\delta(\sqrt{-g} L_m )}{\delta g^{\mu\nu}}.
\end{equation}
\subsection {Solutions on the anisotropic {\bf Universe}}
 The simplest model of the non-isotropic {\bf Universe} is BI model, which exhibit a homogeneity  and special flatness and is the straightforward generalization of the flat FRW. Therefore we introduce the line element of the BI metric as 	
\begin{equation}\label{4}
ds^2=-dt^{2}+A^{2}dx^{2}+B^{2}dy^{2}+C^{2}dz^{2},
\end{equation}
where  the metric functions, $A$, $B$, $C$, are only functions of time, $t$.
 It is  assumed that the matter is perfect fluid, then the energy momentum tensor is given by
\begin{equation}\label{6}
T_{\mu\nu}=\rho u_{\mu}u_{\nu}+ph_{\mu\nu},
\end{equation}
where  $\rho$,  $p$, $h_{\mu\nu}=g_{\mu\nu}+ u_{\mu}u_{\nu}$ and $u^{\mu}$  are the total energy density of a perfect fluid, the corresponding total pressure, the projection tensor and   the flow vector  respectively, and the latter satisfying the following relation
\begin{equation}\label{7}
u_{\mu}u^{\mu}=-1.
\end{equation}
The mean Hubble parameter $H$ is obtained  by
\begin{equation}\label{12}
H=\frac{1}{3}\theta=\frac{\dot{a}}{a},
\end{equation}
where $\theta=u^{j}_{;j}$, $a=(ABC)^{1/3}$  are the  scalar expansion and   the scale factor respectively.  It is defined the  shear scalar  $\sigma_{ij}$ as  \cite{8}
\begin{equation}\label{13}
\sigma_{ij}=u_{i,j}+\frac{1}{2}(u_{i;k}u^{k}u_{j}+u_{j;k}u^{k}u_{i})+\frac{1}{3}\theta h_{ij},
\end{equation}
by using Eqs. (\ref{4})  and (\ref{13}), we obtain
\begin{equation}\label{14}
\sigma^{2}=\frac{1}{2}{\biggr [ (\frac{\dot{A}}{A})^{2}+
(\frac{\dot{B}}{B})^{2}+(\frac{\dot{C}}{C})^{2} \biggl
]}-\frac{1}{6}\theta^{2}.
\end{equation}
 In the context of BI {\bf Universe}, the Ricci scalar is given by
\begin{equation}\label{17}
R=6\big( \dot{H} + 2 H^2\big) + 2 \sigma^2.
\end{equation}
 We assume  the {\bf Universe}  filled with perfect fluids and it satisfies the effective equation of state $\omega = p/\rho$ and $T^{m} = T^{m \mu}_{\mu}$. Combining  Eqs. (\ref{2})-(\ref{6}) gives
\begin{equation}\label{18}
\kappa^{2}\rho=\frac{f(R)}{2}-\left(3(\dot{H}+H^2)+2\sigma^2\right)f'(R)+3H\dot{R}f''(R),
\end{equation}
\begin{eqnarray}\label{19}
\kappa^{2}p&=&-\frac{f(R)}{2}+(3H^2+\dot{H})f'(R)-(2H\dot{R}+\ddot{R})f''(R)-f'''(R)\dot{R}^2.
\end{eqnarray}
Moreover, the Bianchi identities give an another condition on the conservation of the energy
\begin{equation}\label{20}
\dot{\rho}+\theta(\rho+p)=0.
\end{equation}
Solving    Eqs. (\ref{18})-(\ref{19}), requiring  an explicit form  from $f(R)$ gravity. Hence, we take advantage of  a  $f(R)$ function proposed in  Ref. \cite{ali}
 \begin{equation}\label{11a}
 f(R)=R+R\ln {[\frac{R}{R_{c}}]}^{\mp\epsilon},
 \end{equation}
 where $R_c$ is positive constants. It is clear $f(R)\vert_{R=0}=0$,  at the flat space time and  in  the $\epsilon\ll 1 $, Eq. (\ref{11a}) reduced to
 \begin{equation}\label{12}
 f(R)=R({\frac{R}{R_c}})^{\mp\epsilon},
 \end{equation}
or for convenience, one can write
\begin{equation}\label{21}
f(R)=R^{1\pm\epsilon},
\end{equation}
where $\epsilon$ is a constant. For obtain stability of the above  $f(R)$,   the following condition  must   be satisfied \cite{ak}.
\begin{equation}\label{22}
\frac{d^2 f}{dR^2}\geq0.
\end{equation}
 For determining of $+\epsilon$,  by employ  the above constraint  and limit of  power-law solution in the expanding anisotropic {\bf Universe},  the $\epsilon$  form should satisfy the following relations \cite{sn1,snd3}
\begin{equation}\label{23}
\epsilon\geq0.
\end{equation}
In order to gain better insight we assume $A(t)=t^{\alpha}$ and $B(t)=C(t) = t^{n\alpha}$, where $\alpha $ and $ n$ are real constant. {\bf For a positive expand  Universe,  the allowed ranges for the parameter $\alpha$ is  $\alpha\geq 1.1$  and against, small deviations from the isotropic  background  the $n$ parameter is taken $0.9\leq n\leq1.1$}. Furthermore, one can define a scale factor as $a(t) = (ABC)^{1/3} = t^{\alpha(1+2n)/3}$. So based   on this definition,  we obtain the Hubble parameter,  the shear   and   Ricci scalar are given by
\begin{eqnarray}\label{24}
H&=&\frac{\alpha(1+2n)}{3t},\\
\sigma^{2}&=&\frac{\alpha^2(1-n)^2 }{3t^2},\label{24a}\\
R&=& \frac{2\alpha^2(1+2n+3n^2) - 2\alpha(1+2n)}{t^2}. \label{24b}
\end{eqnarray}
 By using Eqs. (\ref{21}), (\ref{24}), (\ref{24a}) and  (\ref{24b}), Eqs. (\ref{18}), (\ref{19}) and (\ref{20}) end up where $\epsilon=\alpha(1+2n)/2-1/2$. From (\ref{21}) it is clearly obvious that for $\alpha=1/(1+2n)$,  the obtained $f(R)$, (\ref{21}), reduces to $f(R)=R$ which is the standard Einstein general relativity action.
From Eq. (\ref{24b}), it follows that
\begin{equation}\label{25}
\dot{R}=-\frac{4\alpha}{t^3}[\alpha(1+2n+3n^2) - (1+2n)].
\end{equation}
If the space time to be isotropic, $\sigma=0$  ($n=1$), Eq. (\ref{25}) reduce to the result of \cite{7}.
\section{Gravitational Baryogenesis  in Anisotropic {\bf Universe}}
 The existent mechanism  in \cite{12} to generate baryon asymmetry is  proposed by
the presentation  a dynamical breaking CPT.  The responsible interaction  is specified by  coupling between the derivative of the Ricci scalar curvature $R$ and the baryon current $j^{\mu}$ \cite{7}.
The Ricci scalar and the baryon number current, $J^{\mu}$, are given by
\begin{equation}\label{28}
S_{int}=\frac{\varepsilon}{M_{*}^{2}}\int d^{4}x
\sqrt{-g}(\partial_{\mu}R)J^{\mu},
\end{equation}
where $M_{*}$ is the cutoff scale characterizing the effective theory and $\varepsilon=\pm1$.
The baryon number density in the thermal equilibrium has been
worked out in detail in \cite{12}. It lead to
\begin{equation}\label{29}
n_{b}=n_{B}-n_{\bar{B}}=\frac{g_{b}T^{3}}{6\pi^{2}}\left(\frac{\pi^{2}\mu_{B}}{T}+(\frac{\mu_{B}}{T})^3\right),
\end{equation}
where $\mu_{B}$ is a chemical potential and
$\mu_{B}=-\mu_{\bar{B}}=-\epsilon\dot{R}/M_{*}^{2}$ and $g_{b}\simeq
1$ is the number of internal degrees of freedom of baryons.
The baryon number to entropy
ratio is \cite{12}
\begin{equation}\label{30}
\frac{n_{b}}{s}\simeq-\epsilon\frac{15g_{b}}{4\pi^{2}g_{s}}\frac{\dot{R}}{M_{*}^{2}T}|_{T_{D}},
\end{equation}
 where $s=(2\pi^{2}/45)g_{s}T^{3}$, where $g_{s}\simeq106$. In an expanding {\bf Universe} the baryon number violation decouples at a temperature denoted by $T_{D}$ and a net baryon asymmetry is remaining.
  \subsection{$f(R)= R$}
  For  the case of $\epsilon=0$ and using Eqs. (\ref{25}) and (\ref{30}),  the baryon asymmetry in terms of temperature can be determined by
\begin{equation}\label{31}
\frac{n_{b}}{S}\simeq\epsilon\frac{15g_{b}}{  \pi^{2}g_{s}}\frac{ \alpha[\alpha(1+2n+3n^2) - (1+2n)]}{M_{*}^{2}T_D t^3_D},
\end{equation}
where $t_D$ is the decoupling time.  An important assumption that needs to be taken into account is that the vacuum decays ``adiabatically", that
means the specific entropy of the massless particles   remains constant irrespective of the total entropy, which may increase. In this case, some equilibrium relations are hold but not all \cite{12a}, for instance the energy density versus temperature, $\rho_R\propto T_R^4$, particle number versus temperature,  $n_R\propto T_R^3$, but   the temperature does not obey the scaling relation  $T_R\propto a^{-1}$. In what follows,  by using the following usual expression of the energy density \cite{12b, 13, 14}  and   equating it to  the expression of $\rho$ is given by Eq. (\ref{18})
\begin{equation}\label{32}
\rho=KT^{4},
\end{equation}
where $K=g_s\pi^2/30$ is proportional to the total number of effectively
degree of freedom, we get
\begin{eqnarray}\label{33}
T=(\frac{30}{\pi^2g_s})^{\frac{1}{4}}\frac{(\alpha^{2}n(n+2))^{\frac{1}{4}}m_p^{\frac{1}{2}}}{t^{\frac{1}{2}}},
\end{eqnarray}
where $m_{p}\simeq1.22\times10^{19}GeV$ is the Planck mass.  Inserting (\ref{33}) into (\ref{31}), one obtain
\begin{eqnarray}\label{34}
\frac{n_{b}}{S}\simeq2.95(\frac{T_D}{m_p})^{5}(\frac{m_p}{M_{*}})^{2} G{(\alpha, n)},
\end{eqnarray}
where $G{(\alpha, n)}=\frac{\alpha(1+2n+3n^2)-(1+2n)}{\alpha^{2}(n(n+2))^{\frac{3}{2}}}$.  Fig. 1,  shows the $\frac{n_b/s}{\gamma}$  versus  $\alpha$ for three different values of the  $n$ parameter. It is clears that the curve is shifted to the smaller value of the  $\frac{n_b/s}{\gamma}$ with increasing  $n$ as can be seen from the diagram, also, the  $\frac{n_b/s}{\gamma}$ decrease  with increasing of the $\alpha$ parameter, in addition, by increasing of the $n$, its slope  become sharper. Fig. 2, illustrate
 the $\frac{n_b/s}{\gamma}$ versus $n$  parameter for three different values of the $\alpha$ parameter. It is clear that  the curve is shifted to the smaller   value of the  $\frac{n_b/s}{\gamma}$ with increasing  $\alpha$,  as can be seen from the diagram, also, the  $\frac{n_b/s}{\gamma}$ decrease with increasing of the $n$ parameter, and its slope become sharper.
\begin{figure}[tb]
 \centerline{\includegraphics[width=.42\textwidth]{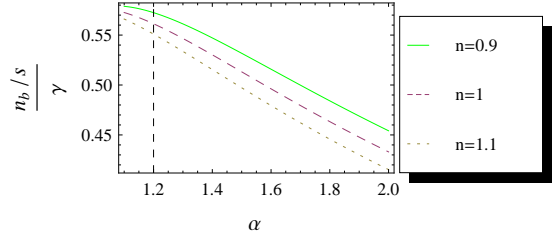}}
\caption{The plot shows the  evolution of the $\frac{n_b/s}{\gamma}$, Eq. (\ref{34}), versus $\alpha$ for three  different values of the $n$ parameter and $\gamma=2.95(\frac{T_D}{m_p})^{5}(\frac{m_p}{M_{*}})^{2}$. }
 \label{fig:1}
 \end{figure}

 \begin{figure}[tb]
 \centerline{\includegraphics[width=.42\textwidth]{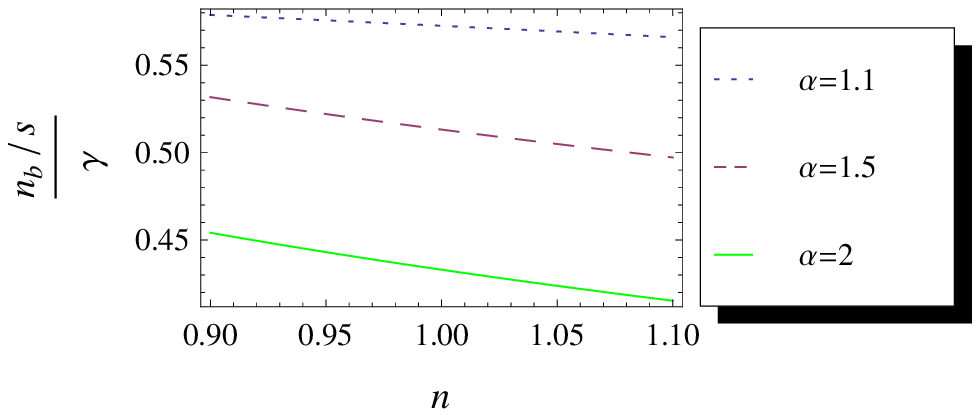}}
\caption{The plot shows the  evolution of the $\frac{n_b/s}{\gamma}$, Eq. (\ref{34}), versus $n$ for three  different values of the $\alpha$ parameter and $\gamma=2.95(\frac{T_D}{m_p})^{5}(\frac{m_p}{M_{*}})^{2}$.  }
 \label{fig:2}
 \end{figure}

\subsection{$f(R) = (\frac{R}{A})^{1+\epsilon}$}
As an example, we examine the case that $f(R) = (\frac{R}{A})^{1+\epsilon}$, where $A=m_p^{2-\frac{2}{1+\epsilon}}$. In this case,  using Eqs. (\ref{18}), (\ref{24}), (\ref{24a}), (\ref{24b}), (\ref{25}) and (\ref{32}) and contribute    the $\epsilon=\alpha(1+2n)/2-1/2$,
we can obtain the decoupling  temperature $T_D$ as a function of the decoupling time $t_D$, namely
\begin{eqnarray}\label{35}
T_D=(\frac{1}{\pi^2g_s})^{\frac{1}{4}} \chi_{\alpha n}^{\frac{1}{4}}\frac{m_p^{\frac{1}{2}}}{(At_D^2)^{\frac{1+\alpha(1+2n)}{8}}},
\end{eqnarray}
where
$$\chi_{\alpha n}=15\alpha(1+2n)\bigg(2\alpha[-1+\alpha+n(-2+(2+3n)\alpha)]\bigg)^{\frac{1+\alpha(1+2n)}{2}}.$$
 Inserting (\ref{35}) into (\ref{30}), one obtain
\begin{eqnarray}\label{36}
\frac{n_{b}}{S}\simeq 0.014 (\pi^2g_s)^{\frac{3}{1+\alpha(1+2n)}}H_{\alpha n}\frac{A^{\frac{3}{2}}T_D^{-1+\frac{3}{1+\alpha(1+2n)}}}{M_{*}^2m_p^{\frac{6}{1+\alpha(1+2n)}}},
\end{eqnarray}
where
$H_{\alpha n}=\alpha[-1+\alpha+n(-2+(2+3n)\alpha)]\chi_{\alpha n}^{-\frac{3}{1+\alpha(1+2n)}}$.
\begin{figure}[tb]
 \centerline{\includegraphics[width=.42\textwidth]{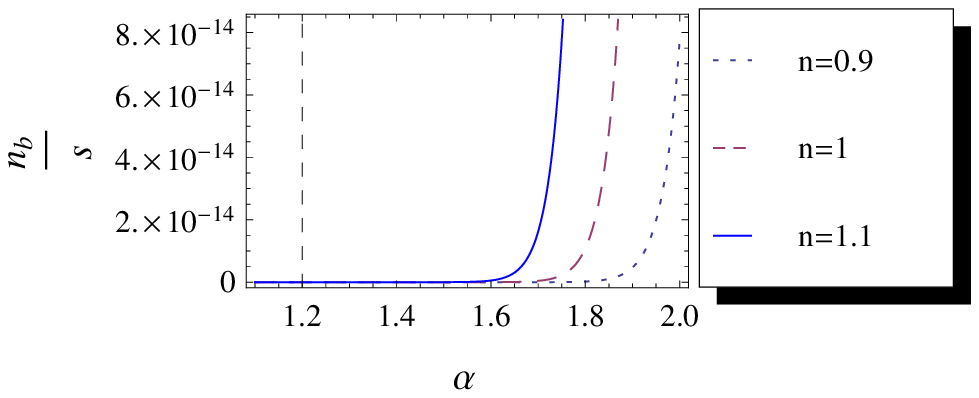}}
\caption{   The plot shows the  evolution of the  $n_b/s$, Eq. (\ref{36}), versus $\alpha$ for three  different values of the $n$ parameter. }
 \label{fig:3}
 \end{figure}
 \begin{figure}[tb]
 \centerline{\includegraphics[width=.42\textwidth]{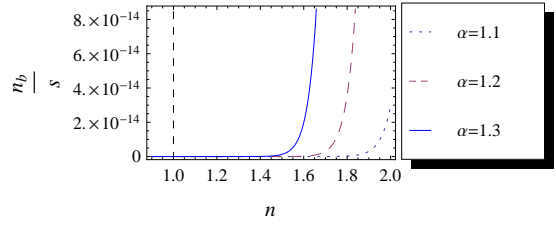}}
\caption{   The plot shows the  evolution of the  $n_b/s$, Eq. (\ref{36}), versus $n$ for three  different values of the $\alpha$ parameter. }
 \label{fig:4}
 \end{figure}
By assuming that the cutoff scale $M_{*}$ takes the value $M_{*}=10^{12} GeV$, also that the critical temperature is equal to   $T_D=2\times 10^{16} GeV$ \cite{7}, and  $m_p=1.22\times 10^{19} GeV$, $g_s=106$, $\alpha=1.8$, and finally for  $n=0.9$, the
baryon-to-entropy ratio becomes  $n_b/s\simeq8.5\times10^{-14}$, which is in very good agreement with observations. Fig. 3, shows the  $n_b/s$  versus  $\alpha$ for three different values of the  $n$ parameter.  It is clears that  the  $n_b/s$ is from order of $10^{-14}$  for  the larger value of the  $\alpha$, against the different values of the $n$ parameter,  but  with  decreasing  $\alpha$, the   $n_b/s$ close to zero value, in addition, by increasing the $n$ parameter, the diagram with sharper slope  is shifted to the smaller value of $\alpha$.
Fig. 4,  shows the  $n_b/s$  versus  $n$ for three different values of the  $\alpha$ parameter. It is clears that  the  $n_b/s$ is  from order of $10^{-14}$  for  the larger value of the  $n$, against the different values of the $\alpha$ parameter, but  with  decreasing  $n$, the   $n_b/s$ close to zero value, in addition, by increasing the $\alpha$ parameter, the diagram with a  bit of sharper slope is shifted to the smaller value of $n$.
  As it can be seen, both the parameters $\alpha$ and $n$ affect the baryon-to-entropy ratio in the same way,
and thus the baryon-to-entropy ratio may be used
to constrain the functional form of the $f(R)$ gravity.

\subsection{$f(R) =R+\frac{1}{H_i}R^2$}
{\bf It would be interesting to analyze another realistic model in $f(R)$ gravity \cite{r3,r4,r5,r6} and compare its results   to the preceding case.
\begin{eqnarray}\label{37}
f(R) =R+\frac{1}{H_i}R^2,
\end{eqnarray}
with $H_i\gg1$. The dynamics presented by this model is found to be in agreement with the data presented by  the  recent observations of the Planck collaboration \cite{r13}.
This model could be helpful in the study of non-singular version of the Starobinsky $R^2$ inflation model.  The same as preceding case, decoupling temperature   for this model becomes
\begin{eqnarray}\label{38}
T=(\frac{ m_p^2\alpha^2}{\pi^2 g_s H_i})^{\frac{1}{4}}\frac{1}{t}\bigg(60[3+\alpha+n(6+(n-2)\alpha)][-1+\alpha+n(-2+(2+3n)\alpha)]-30nH_i t^2(n+2)\bigg)^{\frac{1}{4}}.
\end{eqnarray}
By substituting $T$ from Eq. (\ref{38}) into (\ref{30}), the final expression for the baryon-to-entropy ratio  is equal to
\begin{eqnarray}\label{39}
\frac{n_{b}}{S}\simeq \frac{15}{M^2_*\pi^{\frac{3}{2}}t_D^2}\sqrt{\frac{\alpha}{m_p}}(\frac{H_i}{g_s^3})^{\frac{1}{4}}\frac{-1+\alpha+n(-2+(2+3n)\alpha)}{\bigg(60[3+\alpha+n(6+(n-2)\alpha)][-1+\alpha+n(-2+(2+3n)\alpha)]-30nH_i t^2(n+2)\bigg)^{\frac{1}{4}}},
\end{eqnarray}\\
where $H_i=6.3\times 10^{13}s^{-1}$ \cite{r5,r13}, and $t_D=3\times 10^{-13}s$, for $\alpha=1.1, n=1$.
Fig. 5, shows the  $n_b/s$  versus  $\alpha$ for three different values of the  $n$ parameter.  It is clears that  the  $n_b/s$ is from order of $10^{-11}$, and the   $n_b/s$ is increased by increasing of $\alpha$, in addition, by increasing the $n$ parameter, the diagram is shifted to the larger value of  $n_b/s$.
Fig. 6,  shows the evolution of  $n_b/s$  versus $n$ is the same as Fig. 5. In addition, it is clear that the  $n_b/s$ ratio is in better  agreement with observations rather than the prior case.}
\begin{figure}[tb]
 \centerline{\includegraphics[width=.42\textwidth]{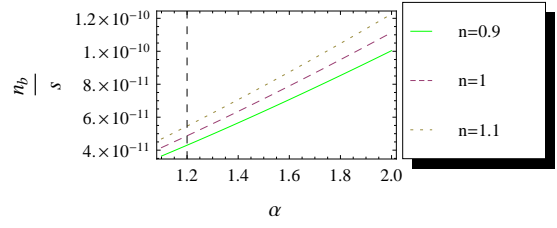}}
\caption{   The plot shows the  evolution of the  $n_b/s$, Eq. (\ref{39}), versus $\alpha$ for three  different values of the $n$ parameter. }
 \label{fig:5}
 \end{figure}
\begin{figure}[tb]
 \centerline{\includegraphics[width=.42\textwidth]{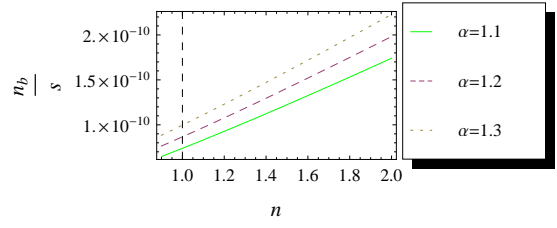}}
\caption{   The plot shows the  evolution of the  $n_b/s$, Eq. (\ref{39}), versus $n$ for three  different values of the $\alpha$ parameter. }
 \label{fig:4}
 \end{figure}


\section{Conclusion}
The main purpose of the present work  has been  the study of the $f(R)$
theory of gravity in an anisotropic metric and evaluate its effect on the baryon number to entropy
ratio.  In the context of $f(R)$ baryogenesis, the baryon-to-entropy ratio
depends on $\dot{R}$, and we discussed three  cases of $f(R)$ theories of gravity, the case $f(R)=R$,  $f(R)=R^{1+\epsilon}$ and also $f(R) =R+\frac{1}{H_i}R^2$.
It has been show that the  $f(R)$ theories of gravity provide a natural  setting in which the baryon asymmetry in the {\bf Universe} may be generated through the mechanism of the gravitational baryogenesis.
 we have shown, the variables $\alpha$ and $n$ plays a crucial role in the calculation of the baryon-to-entropy ratio.
These results depend on the exact solution of $f(R)$ field equations.   Therefore, by select  valid different  $f(R)$ functions, one can examined  the evolution of the baryon number to entropy ratio to obtained  a best agreement to value of $n_{b}/s$ estimated.   Also, we have obtained $\dot{R}$ and $n_{b}/s$ for radiation dominant regime and the
effect of anisotropy of space
 time obviously was seen in it, in addition,  the imprint of shear tensor was inevitable and if fix  $\sigma=0$ then $n_{b}/s=0$. It is clear, that
we have shown  the baryon asymmetry in anisotropic
{\bf Universe} is larger than the baryon asymmetry in Friedmann  Robertson
Walker (FRW) space time.
It is shown that deviations from general relativity on the anisotropic {\bf Universe} in the $f(R)$ gravity  may be considerable in  the study of {\bf Universe. We have considered the plots  $n_{b}/s$ for three  case $f(R)=R$ and $f(R) = (\frac{R}{A})^{1+\epsilon}$, and $f(R) =R+\frac{1}{H_i}R^2$, which,  those results have been explained  in below  the relevant figures. It was explicit, procedure of the evolution  $n_{b}/s$  for the case $f(R)=R$ is subtractive but for $(\frac{R}{A})^{1+\epsilon}$, and $f(R) =R+\frac{1}{H_i}R^2$, this trend is increasing}.
Finally, we close the work drawing a possible connection of this scenario with the extended gravity theories. It is
well known that the physics of the {\bf Universe} at early times constitutes a great laboratory to explore physics beyond the standard cosmological model. For instance, torsion in $f(T)$ gravity can be constrained by big bang
nucleosynthesis \cite{16}. Thus, it is worth noting that the nonzero ratio of baryon-to-entropy of the {\bf Universe} could be a potential quantity to constrain extended
theories of gravity too.

\end{document}